\documentclass[a4paper,twoside,10pt]{letter}
\usepackage{saj,graphicx,multicol,subeqnarray}

\def\papertype{Original Scientific Paper}

\def\msun{$M_{\odot}$}

\def\HII{\hbox{H\,{\sc ii}}}
\newcommand{\SII}{[S\,{\sc ii}]}
\newcommand{\OIII}{[O\,{\sc iii}]}
\newcommand{\Halpha}{H${\alpha}$}
\newcommand{\D}{$^\circ$}
\def\p0{\phantom{0}}
\def\it{\sl}

\def\degr{\hbox{$^\circ$}}
\def\arcmin{\hbox{$^\prime$}}
\def\arcsec{\hbox{$^{\prime\prime}$}}
\def\SNR{\mbox{{SNR~J0536--6735}}}
\def\udc{524.354--77 : 524.722.3}
\begin{document}
\baselineskip=3.1truemm
\columnsep=.5truecm
\newenvironment{lefteqnarray}{\arraycolsep=0pt\begin{eqnarray}}
{\end{eqnarray}\protect\aftergroup\ignorespaces}
\newenvironment{lefteqnarray*}{\arraycolsep=0pt\begin{eqnarray*}}
{\end{eqnarray*}\protect\aftergroup\ignorespaces}
\newenvironment{leftsubeqnarray}{\arraycolsep=0pt\begin{subeqnarray}}
{\end{subeqnarray}\protect\aftergroup\ignorespaces}
%


\markboth{\eightrm MULTIFREQUENCY RADIO OBSERVATIONS OF \SNR}
{\eightrm L.M. Bozzetto, M.D. Filipovi\'c, E.J. Crawford, A.Y. De Horta, M. Stupar}

{\ }

\publ

\type

{\ }


\title{MULTIFREQUENCY RADIO OBSERVATIONS OF \SNR\ (N 59B) with associated pulsar} 


\authors{L.M. Bozzetto$^1$, M.D.~Filipovi\'c$^1$, E.J. Crawford$^1$, A.Y. De Horta$^1$, M. Stupar$^{2,3}$}

\vskip3mm


\address{$^1$School of Computing and Mathematics, University of Western Sydney
\break Locked Bag 1797, Penrith South DC, NSW 1797, Australia}  
\address{$^2$Department of Physics, Macquarie University, Sydney, NSW 2109, Australia}
\address{$^3$Australian Astronomical Observatory, PO Box 296, Epping, NSW 1710, Australia}

\Email{m.filipovic@uws.edu.au}


\dates{April 2012}{April 2012}


\summary{We present a study of new Australian Telescope Compact Array (ATCA) observations of supernova remnant, \SNR. This remnant appears to follow a shell morphology with a diameter of {\it D}=36$\times$29~pc (with 1~pc uncertainty in each direction). There is an embedded \HII\ region on the northern limb of the remnant which made various analysis and measurements (such as flux density, spectral index and polarisation) difficult. The radio-continuum emission followed the same structure as the optical emission, allowing for extent and flux density estimates at 20~cm. We estimate a surface brightness for the SNR at 1~GHz of 2.55$\times$10$^{-21}$ Wm$^{-2}$ Hz$^{-1}$ sr$^{-1}$. Also, we detect a distinctive radio-continuum point source which confirms the previous suggestion of this remnant being associated with a pulsar wind nebulae (PWN). The tail of this remnant isn't seen in the radio-continuum images and is only seen in the optical and X-ray images.
}



\begin{multicols}{2}
{


\section{1. INTRODUCTION}

The Large Magellanic Cloud (LMC) is an irregular dwarf galaxy located at a distance of 50~kpc (Macri et al. 2006). It is considered to be a near ideal galaxy for achieving detailed observations of celestial object such as Supernova Remnants (SNRs). The LMC is located in the direction towards the South Pole (one of the coldest areas of the radio sky,) minimising interference from galactic foreground radiation. Furthermore, the LMC resides outside of the Galactic plane, rendering the influence of dust, gas and stars as negligible.

Predominately non-thermal emission is a well-known characteristic of SNRs in the radio-continuum. SNRs have a typical radio spectral index of $\alpha\sim-0.5$ defined by $S\propto\nu^\alpha$. However, this can significantly vary, as there exists a wide variety of SNRs in differing environments (Filipovi\'c et al.~1998). The morphology, structure, behaviour and evolution of the ISM can be attributed to SNRs, and in turn the ISM heavily impacts the properties of SNRs, as their expansion and evolution are heavily dependant on their surrounding environment.

Type II supernova are the result of core collapse SNRs in large stars with an initial mass greater than 8 $\pm$ 1~\msun (Smartt 2009). Depending on how massive the progenitor star is, the explosion may leave behind a compact central object such as a neutron star, or if spinning in our line of sight, a pulsar. There appears to be a lack of pulsar detections in the MC's when compared to the $\sim$1.79 $\times$ 10$^4$ predicated pulsars as modelled by Ridley \& Lorimer (2010). Out of the 56 confirmed and some 20 candidate LMC SNRs, there are currently only 4 known SNR-pulsar associations (N~49, 30~Dor~B, B0540-693 \& B0453-68), with an additional two candidates, J0529-6653 (Bozzetto et al. 2012) and J0541.8-6659 (Grondin et al. 2012). In contrast, the Milky Way (MW) contains 274 SNRs (Green 2009) and some $\sim$1900 pulsars. { Globally, this lack of larger SNR--pulsar associations in the LMC and MW can be explained by the fact that radio pulsars live a significantly longer life in comparison to their associated SNRs, resulting in them ejecting energy into the ISM long after their SNR has dissipated into the ambient ISM. Also, it maybe be attributed to the fact that many neutrons stars posses different properties to those of conventional radio pulsars (Gotthelp \& Vasisht 2000).}

Davies et al. (1976) described this region as being very bright with a diameter of 8\arcmin$\times$8\arcmin\ and commented about the appearance of optical knots. Clarke et al. (1976) observed this object as a part of their catalogue of radio sources and recorded a peak intensity of 0.36~Jy (408~MHz) and made note of the extended emission. Mathewson et al. (1985) recorded an optical size of 50~pc, integrated flux density of 0.244~Jy (843~MHz) and used this new flux measurement with the 408~MHz measurement (Clarke et al. 1976) to produce a spectral index of $\alpha$=--0.6. They recorded a [S\,{\sc ii}]-to-H$\alpha$ ratio of 0.6 which supported SNR identification. They also noted the strong H{\sc ii} region at the northern shell of the remnant. This object was also observed by Filipovi\'c et al. (1995) in their survey of the Magellanic Clouds, reporting a integrated flux measurement of 0.1096~Jy at 8550~MHz. Haberl \& Piestch (1999) measured an X-ray extent of 56.6\arcsec\ and named this object HP~551. \SNR\ was not detected with the Far Ultraviolet Spectroscopic Explorer (FUSE) in the Blair et al. (2006) survey of the Magellanic Clouds SNRs. Bamba et al. (2006) observed this object with the XMM-Newton, reporting elongated structure with a compact central source. They suggest that this compact source may be a pulsar wind nebulae (PWN) and that the progenitor of this SNR would be a massive star, greater than 20~$M_{\odot}$. Payne et al. (2008) measured optical spectra lines and found that canonical [SII]/\Halpha\ ratio is 0.4. Desai et al. (2010) reported an extent of 2.4\arcmin\ with detection of a young stellar object (YSO) and detection in the NANTEN survey. They make note, however, that this SNR is superimposed on an OB association and therefore uncertain whether the YSO is related to this SNR and it's progenitor.  

Here, we report on new radio-continuum observations of \SNR. The observations, data reduction and imaging techniques are described in Section~2. The astrophysical interpretation of newly obtained moderate-resolution total intensity in combination with the existing Magellanic Cloud Emission Line Survey (MCELS) images are discussed in Section~3.

\section{2. OBSERVATIONS}

We observed \SNR\ with the Australia Telescope Compact Array (ATCA) on the 15$^\mathrm{th}$ and 16$^\mathrm{th}$ of November 2011, using the new Compact Array Broadband Backend (CABB) with an array configuration of EW367 at wavelengths of 3 and 6~cm ($\nu$=9000 and 5500~MHz). Baselines formed with the $6^\mathrm{th}$ ATCA antenna were excluded, as the other five antennas were arranged in a compact configuration. The observations were carried out in the so called ``snap-shot'' mode, totalling $\sim$50 minutes of integration over a 14 hour period. Source PKS~B1934-638 was used for primary calibration and source PKS~B0530-727 was used for secondary (phase) calibration. The \textsc{miriad} (Sault et al.~1995) and \textsc{karma} (Gooch~2006) software packages were used for reduction and analysis. More information on the observing procedure and other sources observed in this session/project can be found in Boji\v{c}i\'c~et~al.~(2007), Crawford~et~al.~(2008a,b; 2010), \v{C}ajko~et~al.~(2009), Bozzetto et al.~(2010; 2012a,b) and de Horta et al. (2012).

{ The 20/13~cm images as well as our 6/3~cm images} were formed using \textsc{miriad} multi-frequency synthesis (Sault and Wieringa~1994) and natural weighting. They were deconvolved using the {\sc mfclean} and {\sc restor} algorithms with primary beam correction applied using the {\sc linmos} task. The 6~cm image has a resolution of 46.4\arcsec$\times$43.0\arcsec\ at PA=0\D\ and an estimated r.m.s. noise of 0.2~mJy/beam. { Our 6/3~cm images encountered a dynamic range problem due to the strong \HII\ region located towards the Northern limb of the remnant, which in turn drowned out the SNR.}

Other observations used in this project included a 36~cm unpublished image (Fig.~1) taken by the Molonglo Synthesis Telescope (MOST) as described by Mills et al. (1984). The 20 \& 13~cm images (Fig.~1) were created from project C354 which made use of 1.5D, 1.5B \& 6C baselines. The 6 \& 3~cm (Fig.~1) images were taken from project C918 (Dickel et al. 2005).

Archival observations at X-Ray wavelengths taken by the XMM-Newton were also used in this project. We also used the Magellanic Cloud Emission Line Survey (MCELS) that was carried out with the 0.6~m University of Michigan/CTIO Curtis Schmidt telescope, equipped with a SITE $2048 \times 2048$\ CCD, which gave a field of 1.35\degr\ at a scale of 2.4\arcsec\,pixel$^{-1}$. Both the LMC and SMC were mapped in narrow bands corresponding to \Halpha, \OIII\ ($\lambda$=5007\,\AA), and \SII\ ($\lambda$=6716,\,6731\,\AA), plus matched red and green continuum bands that are used primarily to subtract most of the stars from the images to reveal the full extent of the faint diffuse emission. All the data has been flux-calibrated and assembled into mosaic images, a small section of which is shown in Fig.~2 \& 3. Further details regarding the MCELS are given by Smith et al. (2006) and at http://www.ctio.noao.edu/mcels. Here, for the first time, we present optical images of this object in combination with our new radio-continuum data.

\section{3. RESULTS AND DISCUSSION}

The remnant \SNR\ displays some distinctive elements of a shell morphology at radio and optical frequencies. While tt is difficult to determine where the centre of the SNR is exactly located, instead, we note the position of the strong point-like source at RA(J2000)=5$^h$36$^m$00.0$^s$, DEC(J2000)=--67\degr35\arcmin09.1\arcsec. There is an embedded \HII\ region at the northern side of the remnant and an even larger/stronger \HII\ region just outside the western region of the remnant. We estimate a diameter of 148\arcsec$\times$120\arcsec$\pm$4\arcsec\ (36$\times$29$\pm$1~pc) for this remnant based on the optical and X-Ray emission, of which followed the radio emission quite closely at the unimpaired southern regions of the SNR. 

At lower radio frequencies (843, 1400 \& 2400~MHz), \SNR\ has a clear association with the central point source discovered initially at X-rays (Bamba et al. 2006). We note that on either side of this SNR, there appears to be significantly less emission, which would be expected of an object (like a pulsar) that is injecting high amounts of energy into the surrounding environment, thus clearing a cone-like path within the confines of the remnant. Bamba et al. (2006) inferred that PWN may be associated with this SNR. If this is a valid connection between the central object and the remnant, this would be the 5$^{th}$ confirmed SNR/PWN association in the LMC (after N49, 30~Dor~B and B0540--693, B0453--68) with two other candidates J0529-6653 (Bozzetto et al. 2012) and J0541.8-6659 (Grondin et al. 2012).

The X-Ray emission for this remnant somewhat differs from the radio and optical appearance. While most of the known SNRs (including SNRs associated with a PWN) exhibit typical shell-like emission, this object has a ``tail'' like structure emanating out from the North-West of the central remnant source. There is a distinct area of peak emission across the SNR located in the `head' of the remnant.

The MCELS image (Fig.~1 \& 2) also shows the point source that is seen at X-Ray and lower radio frequencies. The remnant appears to be predominately \SII\ and \Halpha\ dominated with lower levels of \OIII\ emission. There is a clear distinction between the head and the tail of the emission in which we see the head with significantly stronger emission and the tail dissipating the further away from the head it resides. This would infer that the pulsar of \SNR\ has travelled over a quite substantial distance, especially in comparison to the outer borders of the remnant. Alternatively, this tail may be indicative of emission being blown away from this region, possibly by the PWN.

Measuring integrated flux density for the remnant was difficult as a consequence of the embedded \HII\ region at the northern side of the remnant. Integrated flux density measurements were taken of the point source in this SNR at frequencies of 1400 \& 2400~MHz and used to estimate the spectra of this source (Table~1) of $\alpha$=0.38$\pm$0.37 (Fig.~3). This rather inverted/flat spectra is typical for the PWN (Haberl et al. 2012). A spectral map was created, however the interference from this object influenced the fluxes for the SNR to an extent where measurements were completely unreliable. Because of this, estimating a surface brightness with the current spectra would result in skewed values. Instead, we adopt the typical SNR spectral index of $\alpha$=--0.55 and use this coupled with a 20~cm flux estimate to produce a surface brightness at 1~GHz of 2.55$\times$10$^{-21}$~W~m$^{-2}$ Hz$^{-1}$~sr$^{-1}$. A representation of this surface brightness vs. diameter can be seen in Fig.~4, using the values ({\it D},~$\Sigma$)=(32.6~pc, 2.55$\times$10$^{-21}$ Wm$^{-2}$ Hz$^{-1}$ sr$^{-1}$) for this remnant. It's apparent that this estimate of $\Sigma$--{\it D} falls within the same range of surface brightness-diameter measurements previously taken of LMC SNRs. 

\vskip5mm

\centerline{{\bf Table 1.} Flux Density of Pulsar J0536-6735.}
\vskip2mm
\centerline{
\begin{tabular}{cccccl}
\hline
$\nu$ & $\lambda$ & R.M.S  & Beam Size  & S$_\mathrm{Total}$ \\
(MHz) & (cm)      & (mJy) & (\arcsec) & (mJy) &\\
\hline
1400 & 20 & 0.50 & 13.9$\times$13.2 & 4.20 \\
2400 & 13  & 0.22 & 8.2$\times$7.7 & 5.16 \\
\hline
\end{tabular}}
\vspace{0.5cm}

We did not detect any polarisation associated with this object in either the {\it Q} or {\it U} intensity maps. However, there is a strong source just outside the western field of this SNR as seen in the optical images (Fig.~2) which exhibited strong polarisation. This may have obstructed any polarisation this SNR had, so we can not completely rule out that this source may be significantly polarised.

}

\end{multicols}

\clearpage

\centerline{\includegraphics[trim=40 0 0 0,angle=-90,width=.5\textwidth]{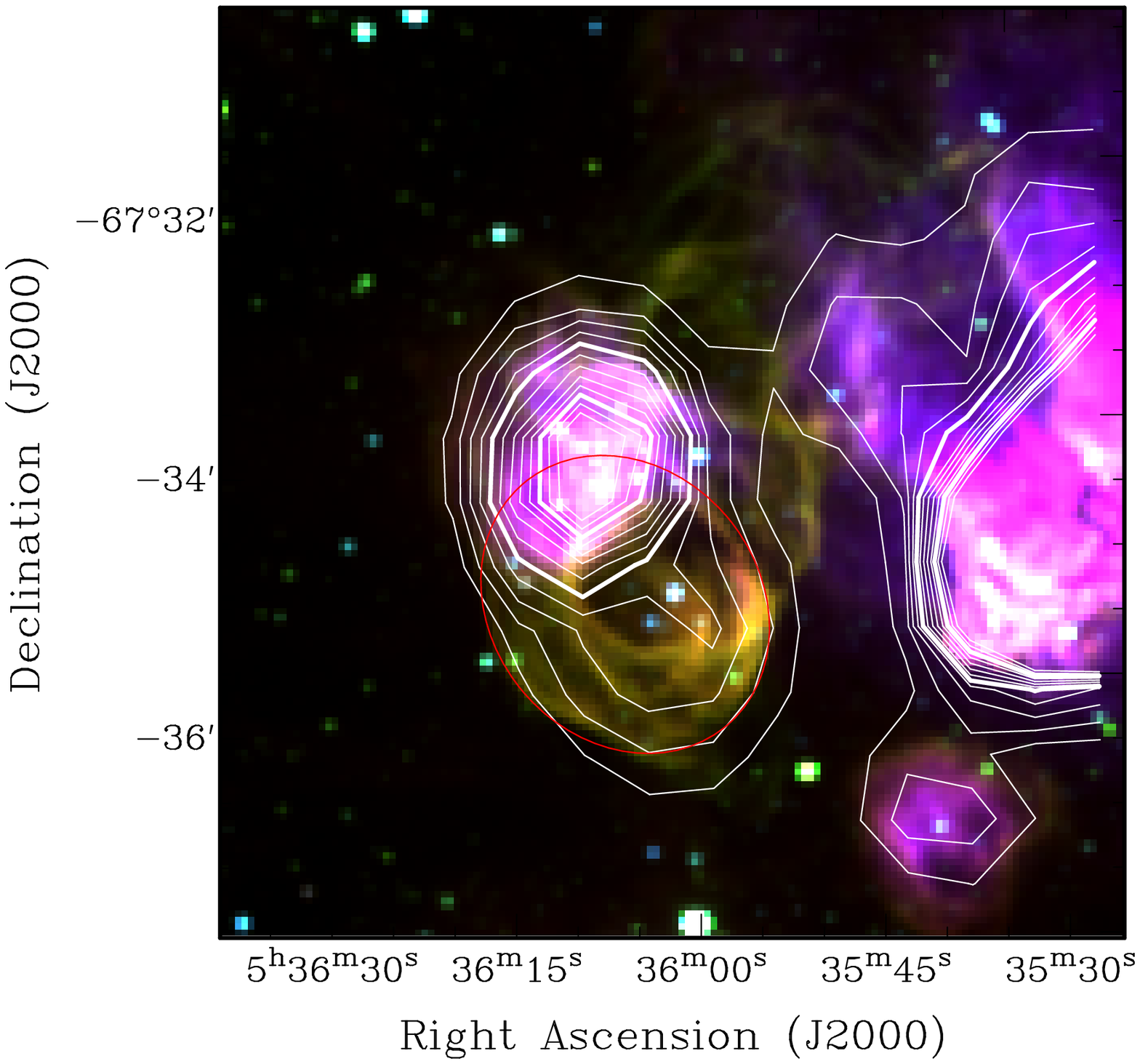}\includegraphics[trim=40 0 0 0,angle=-90,width=.5\textwidth]{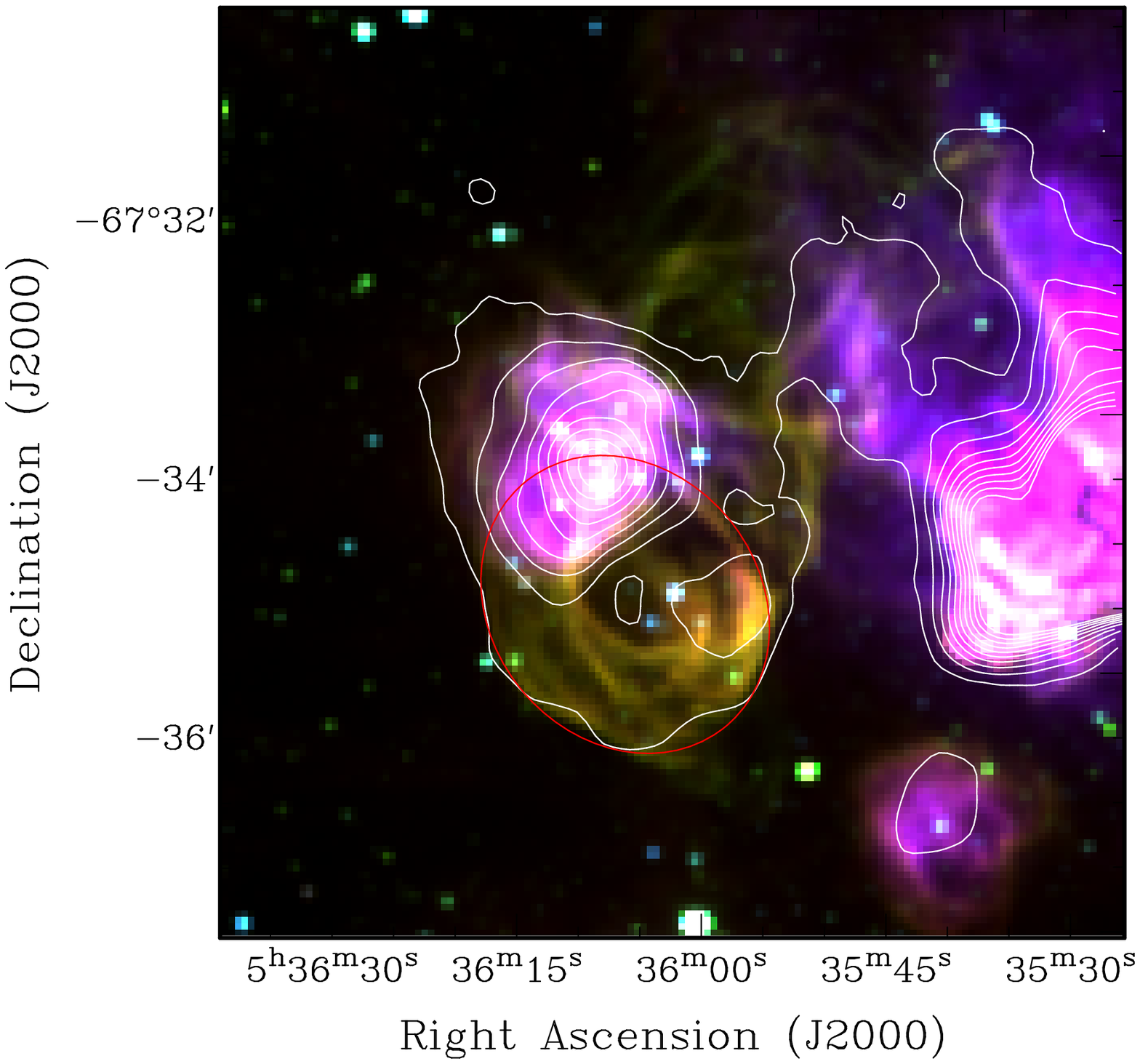}}
\centerline{\includegraphics[trim=60 0 0 0,angle=-90,width=.5\textwidth]{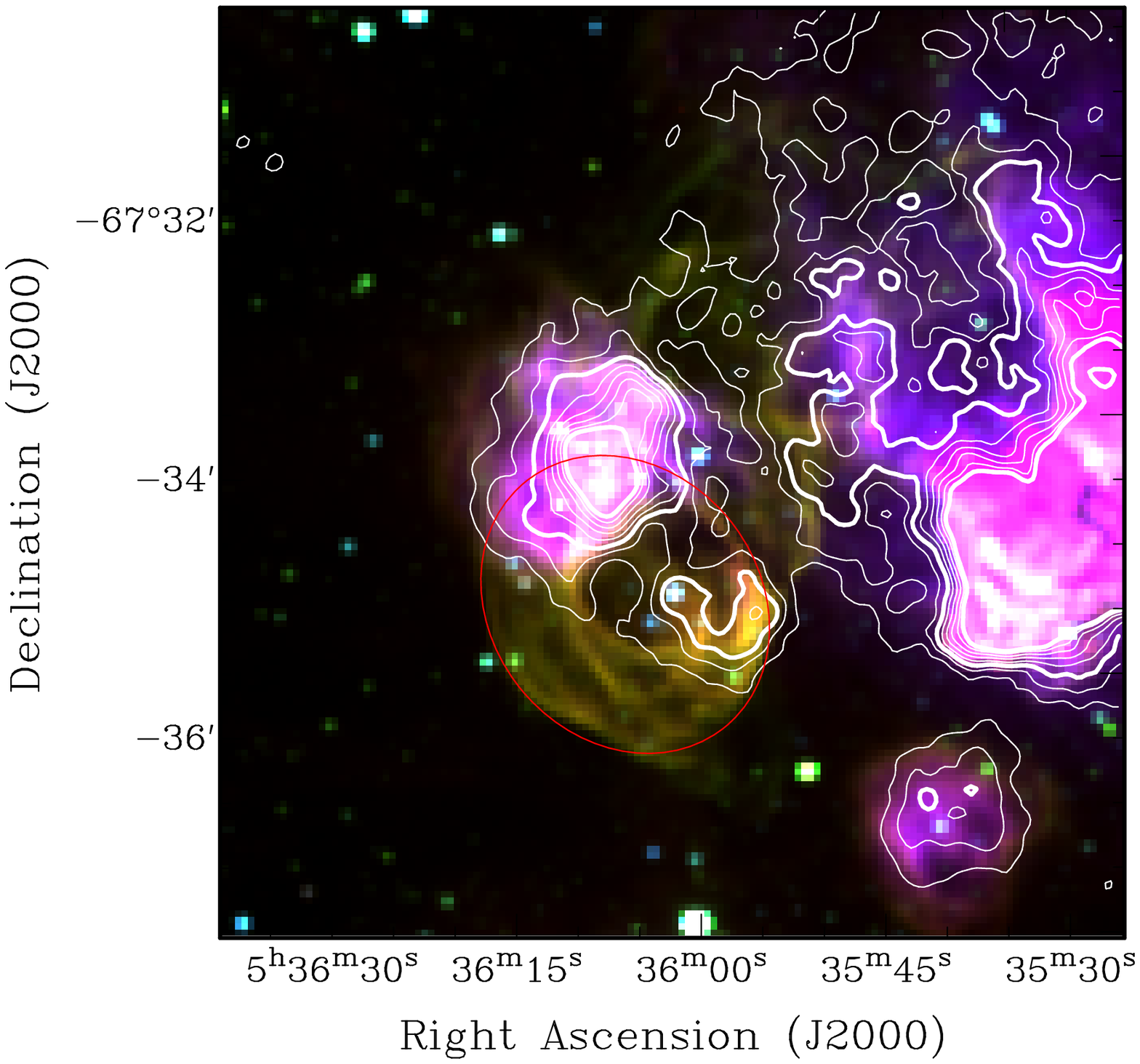}\includegraphics[trim=60 0 0 0,angle=-90,width=.5\textwidth]{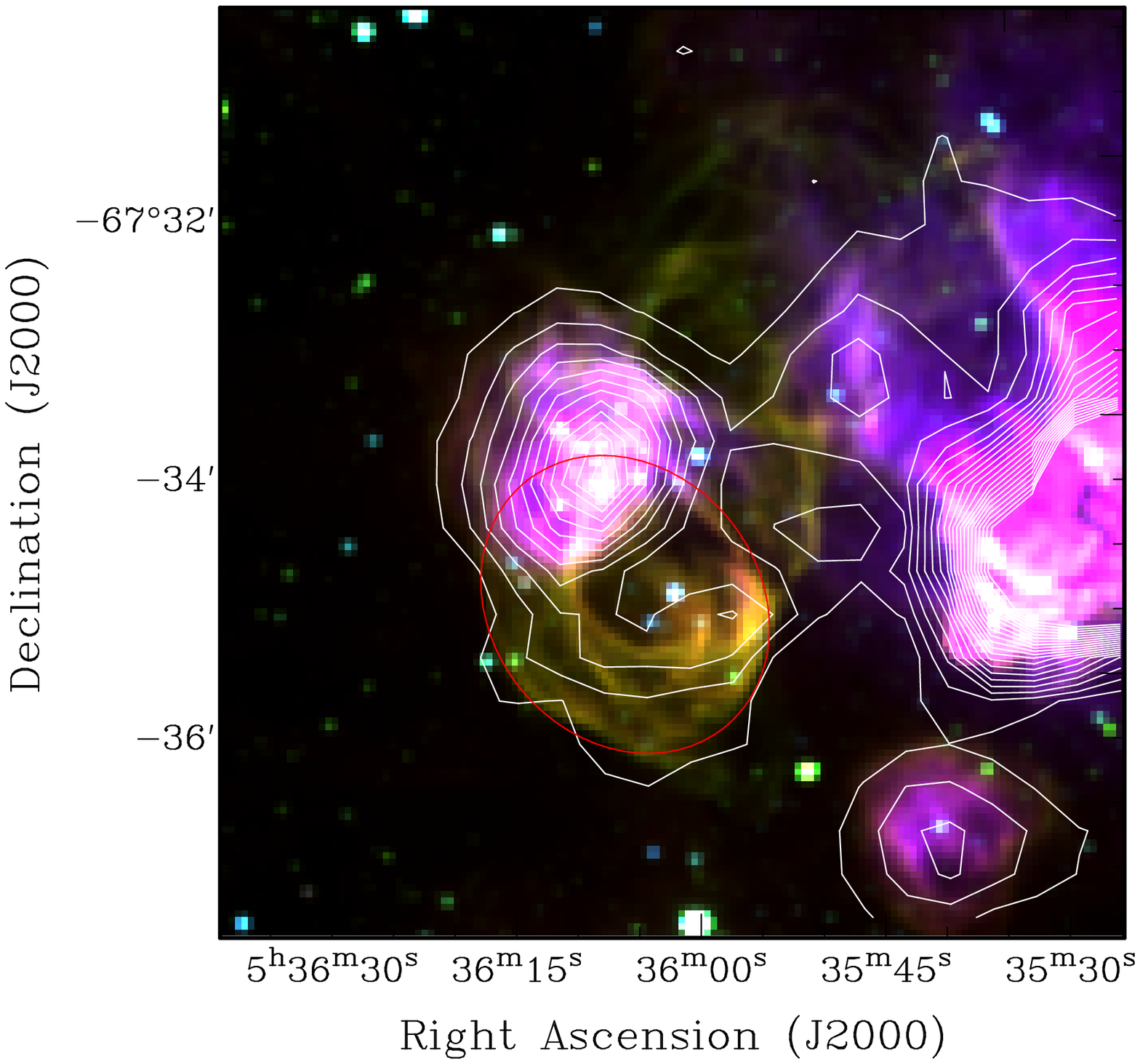}}
\centerline{\includegraphics[trim=60 0 60 0,angle=-90,width=.5\textwidth]{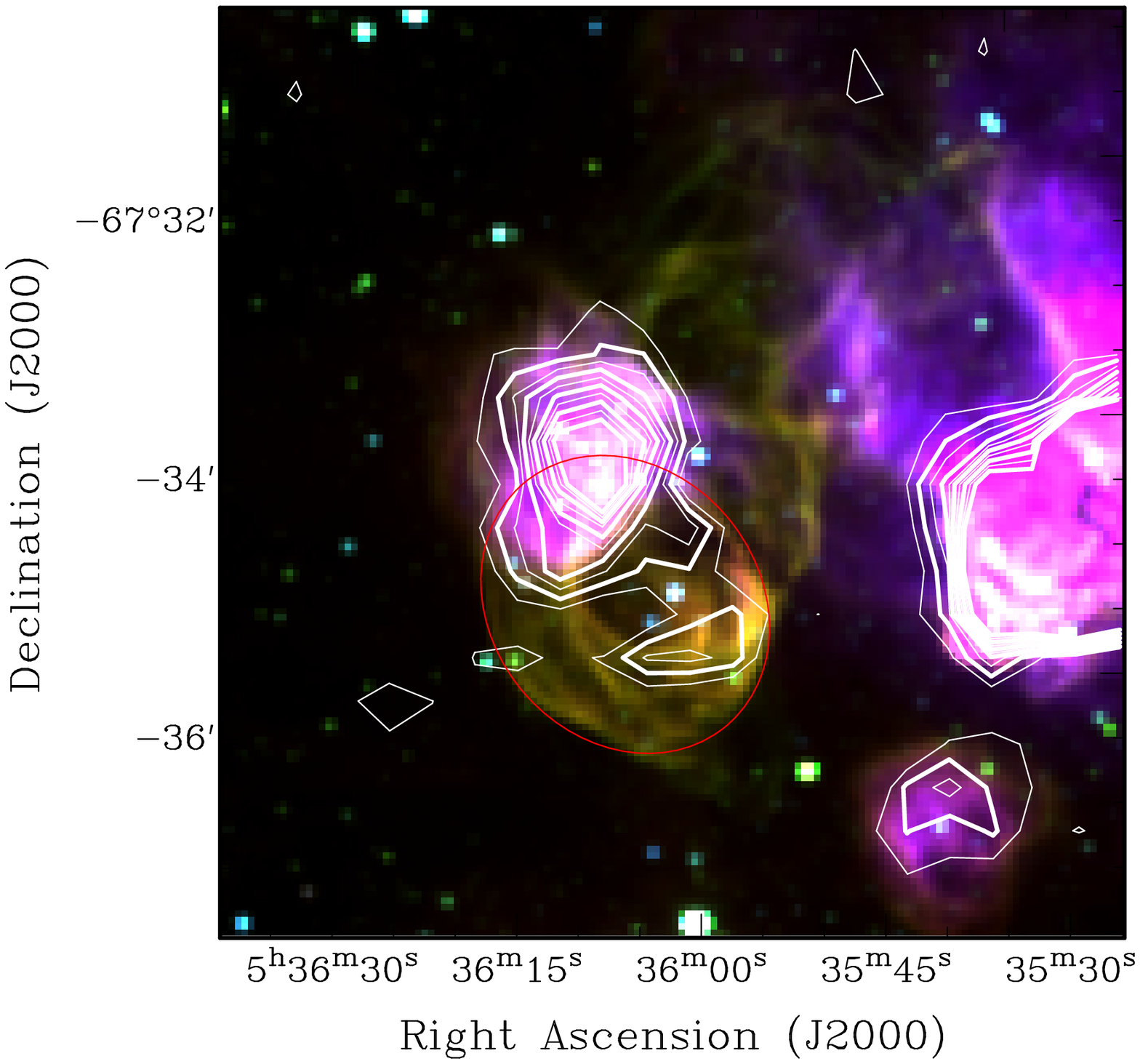}}
\figurecaption{1.}{MCELS composite optical image \textrm{(RGB =H$\alpha$,[S\textsc{ii}],[O\textsc{iii}])}  of \SNR\ overlaid with 36~cm [top left] 20~cm [top right], 13~cm [mid left], 6~cm [mid right] \& 3~cm [bottom] contours.}

\centerline{\includegraphics[trim=0 0 0 0,angle=-90,width=.8\textwidth]{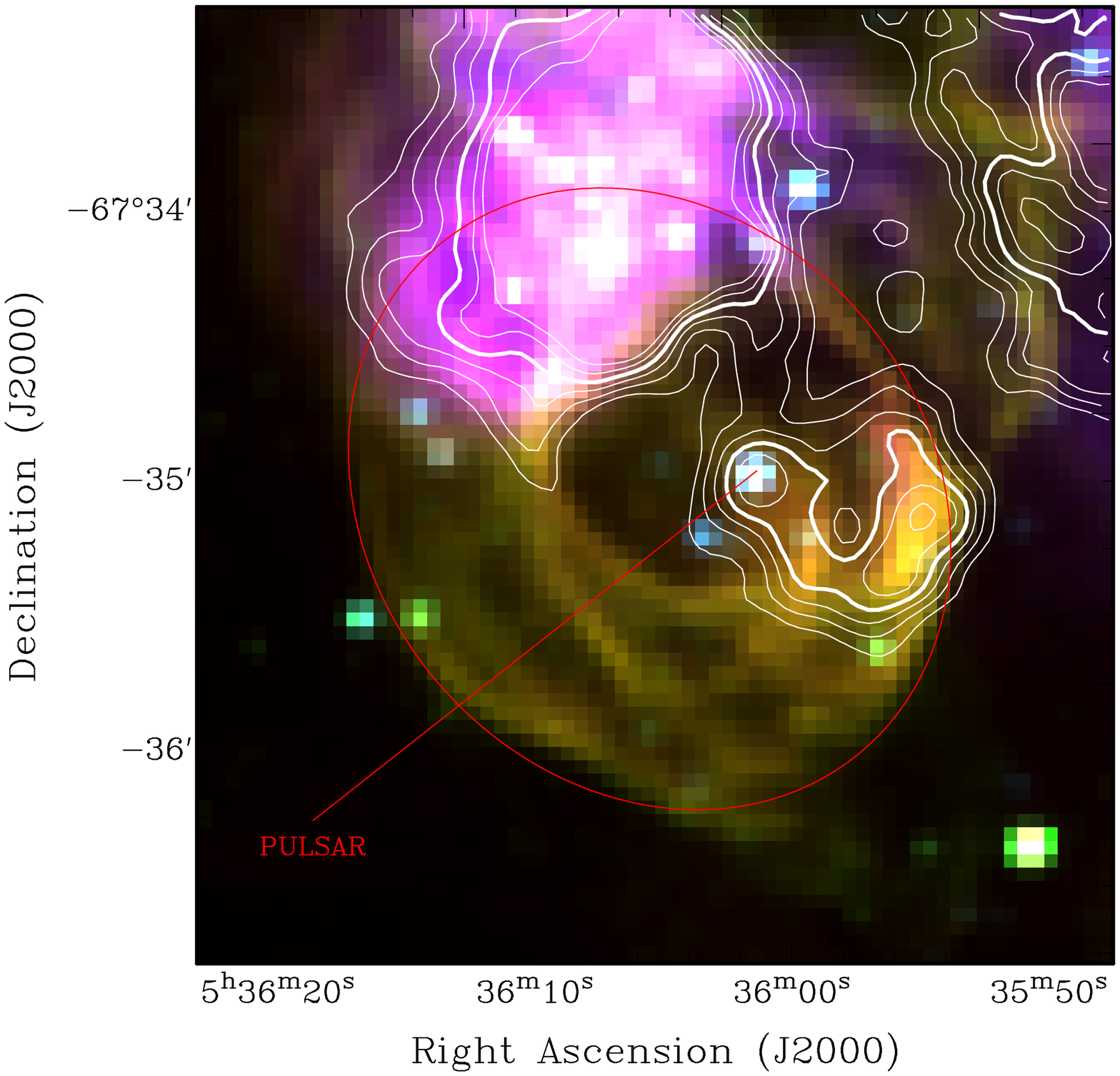}}
\figurecaption{2.}{Close up MCELS image of \SNR\ with 13~cm as seen in Fig.~1. An overlaid annotation file has been added to show the position of the pulsar.}

\centerline{\includegraphics[trim=0 0 40 0,angle=-90,width=.8\textwidth]{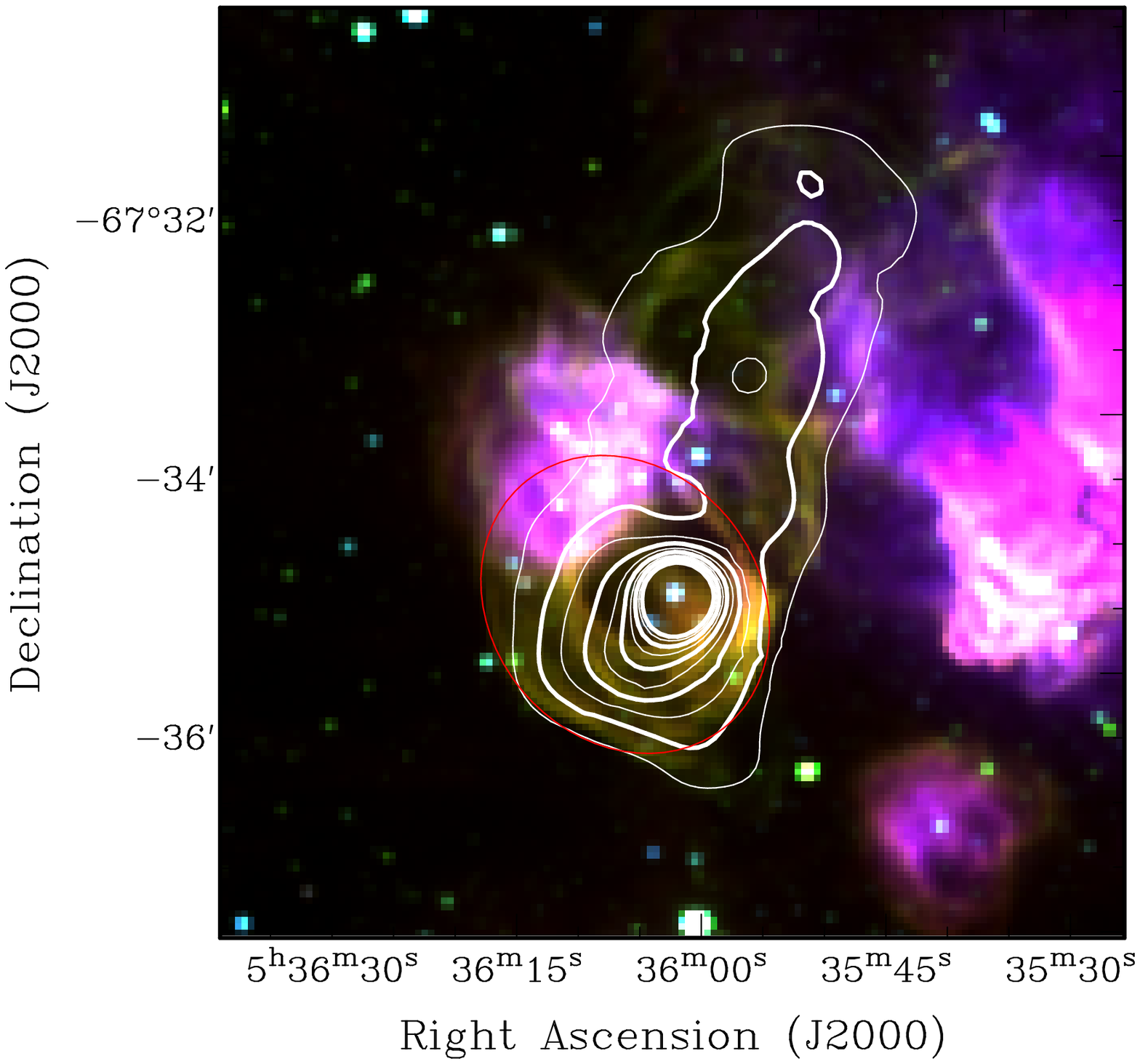}}
\figurecaption{3.}{MCELS composite optical image \textrm{(RGB =H$\alpha$,[S\textsc{ii}],[O\textsc{iii}])} of \SNR\ overlaid with XMM contours.}

\centerline{\includegraphics[trim=0 270 50 150,width=1.5\textwidth]{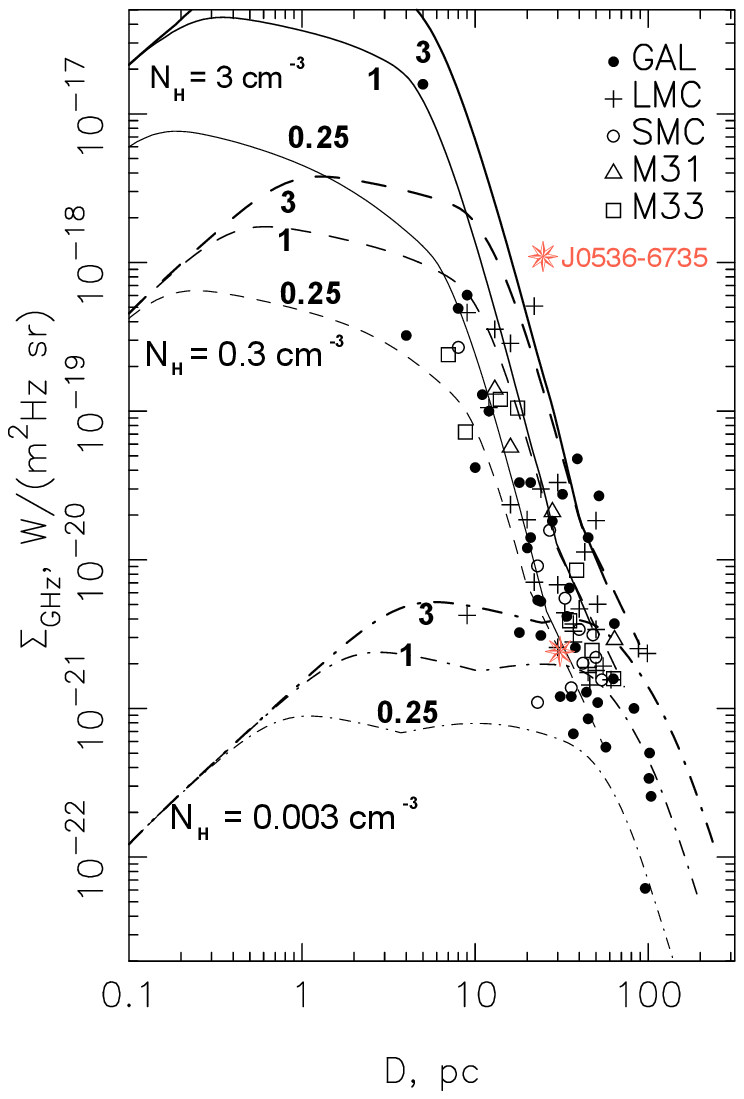}}
\figurecaption{4.}{Surface brightness-diameter diagram from Berezhko \& Volk (2004), with \SNR\ added. The evolutionary tracks are for ISM densities of {\it N}$_H$ = 3, 0.3 and 0.003~cm$^{-3}$ and explosion energies of {\it E}$_{\sc SN}$ = 0.25, 1 and 3 $\times$ 10$^{51}$ erg.}

\begin{multicols}{2}

{

\section{4. CONCLUSION}

\vskip-1mm

This remnant appears to exhibit a shell morphology with an extent of D=(36$\times$29)$\pm$1~pc, a close association between optical and X-Ray images for both the head and tail of the emission, with the radio-continuum images displaying emission from the head of the remnant but none from the tail. Analysis in this paper support previous suggestions of the point source within the head of the SNR being a PWN associated with this remnant.


\acknowledgements{We used the {\sc karma} and {\sc miriad} software package developed by the ATNF. The Australia Telescope Compact Array is part of the Australia Telescope which is funded by the Commonwealth of Australia for operation as a National Facility managed by CSIRO. We thank the Magellanic Clouds Emission Line Survey (MCELS) team for access to the optical images. The MCELS project has been supported in part by NSF grants AST-9540747 and AST-0307613, and through the generous support of the Dean B. McLaughlin Fund at the University of Michigan, a bequest from the family of Dr. Dean B. McLaughlin in memory of his lasting impact on Astronomy.}



\references

\newcommand{\MNRAS}{\journal{Mon. Not. R. Astron. Soc.}}
\newcommand{\ApJ}{\journal{Astrophys. J.}}
\newcommand{\ApJS}{\journal{Astrophys. J. Supplement}}
\newcommand{\AJ}{\journal{Astronomical. J.}}

\vskip-3mm

Bamba, A.; Ueno, M.; Nakajima, H.; Mori, K.; Koyama, K., 2006,  \journal{Astron. Astrophys.}, \vol{450}, 585.
 
Berezhko E. G., Volk H. J., 2004, \journal{Astron. Astrophys.}, \vol{427}, 525.

Blair, W.~P., Ghavamian, P., Sankrit, R., Danforth, C.~W.: 2006, \ApJS, \vol{165}, 480. 

Boji{\v c}i{\'c}, I.~S., Filipovi{\'c}, M.~D., Parker, Q.~A., Payne, J.~L., Jones, P.~A., Reid, W., Kawamura, A., Fukui, Y.: 2007, \MNRAS, \vol{378}, 1237.

Bozzetto, L.~M., Filipovi{\'c}, M.~D., Crawford, E.~J., Boji{\v c}i{\'c}, I.~S., Payne, J.~L., Mendik, A., Wardlaw, B. and de Horta, A.~Y.,\ 2010, \journal{Serb. Astron. J.}, \vol{181}, 43.

Bozzetto, L.~M., Filipovi{\'c}, M.~D., Crawford, E.~J., Payne, J.~L., De Horta, A.~Y. and Stupar, M.,\ 2012, \vol{48}, 41

Bozzetto, L.~M., Filipovi{\'c}, M.~D., Crawford, E.~J., Haberl, F., Sasaki, M., Uro{\v s}evi{\'c}, D., Pietsch, W., Payne, J.~L., de Horta, A.~Y., Stupar, M., Tothill, N.~F.~H., Dickel, J., Chu, Y.-H. and Gruendl, R.,\ 2012, MNRAS, 420, 2588.

\v{C}ajko K.~O., Crawford E.~J., Filipovi{\'c}, M.~D.: 2009, \journal{Serb. Astron. J.}, \vol{179}, 55. 


Clarke, J.~N., Little, A.~G., Mills, B.~Y.: 1976, \journal{Aust. J. Phys. Astrophys. Suppl.}, \vol{40}, 1. 

Crawford, E.~J., Filipovi{\'c}, M.~D. and Payne, J.~L.: 2008a, \journal{Serb. Astron. J.}, \vol{176}, 59. 

Crawford, E.~J., Filipovi{\'c}, M.~D., De Horta, A.~Y., Stootman, F.~H., Payne J.~L.: 2008b, \journal{Serb. Astron. J.}, \vol{177}, 61.

Crawford, E.~J., Filipovi{\'c}, M.~D., Haberl, F., Pietsch, W., Payne, J.~L., De Horta, A.~Y.: 2010, \journal{Astron. Astrophys.}, \vol{518}, A35. 

Davies, R.D., Elliott, K. H., Meaburn, J.: 1976, \journal{Mon. Mem. Royal Astron. Society}, \vol{81}, 89.

Desai, K.~M., Chu, Y.-H., Gruendl, R.~A., Dluger, W., Katz, M., Wong, T, Chen, C.-H.~R., Looney, L.~W., Hughes, A., Muller, E., Ott, J. \& Pineda, J.~L., 2010, \journal{The Astronomical Journal}, \vol{140}, 584.

de Horta, A. Y., Filipovi{\'c}, M.~D., Bozzetto, L. M., Maggi, P., Haberl, F., Crawford, E. J., Sasaki, M., Uro{\v s}evi{\'c}, D., Pietsch, W., Gruendl, R., Dickel, J., Tothill, N. F. H., Chu, Y.-H., Payne, J. L., Collier, J. D.: 2012, \journal{Astron. Astrophys.}, \vol{540}, A25.

Dickel, J. R.; McIntyre, V. J.; Gruendl, R. A.; Milne, D. K., 2005, \journal{The Astronomical Journal}, \vol{129}, 790.

Filipovi\'c, M.~D., Haynes, R.~F., White, G.~L., Jones, P.~A., Klein, U., Wielebinski, R.: 1995, \journal{Astron. Astrophys. Suppl. Series}, \vol{111}, 331.

Filipovi\'c, M.~D., Haynes, R.~F., White, G.~L., Jones, P.~A.: 1998, \journal{Astron. Astrophys. Suppl. Series}, \vol{130}, 421.

Gooch, R.: 1996, in "Astronomical Society of the Pacific Conference Series, Vol. 101, Astronomical Data Analysis Software and Systems V", G. H. Jacoby \& J. Barnes, ed., 80.

Gotthelf E. V., Vasisht G., 2000, in Kramer M., Wex N., Wielebinski R., eds,
IAU Colloq. 177, ASP Conf. Ser. Vol. 202, Pulsar Astronomy - 2000
and Beyond. Astron. Soc. Pac., San Francisco, p. 699.

Green D. A., 2009, Bull. Astron. Soc. India, \vol{37}, 45.

Grondin, M.-H.; Sasaki, M.; Haberl, F.; Pietsch, W.; Crawford, E. J.; Filipovi\'c, M. D.; Bozzetto, L. M.; Points, S.; Smith, R. C., 2012,  \journal{Astron. Astrophys.}, \vol{539}, 15.

Haberl, F., Pietsch, W.: 1999, \journal{Astron. Astrophys. Suppl. Series}. \vol{139}, 277.

Macri, L. M., Stanek, K. Z., Bersier, D., Greenhill, L. J. \& Reid, M. J., 2006, \ApJ, \vol{652}, 1133.

Mathewson, D. S.; Ford, V. L.; Tuohy, I. R.; Mills, B. Y.; Turtle, A. J.; Helfand, D. J., 1985 \journal{Astrophysical Journal Suppl. Series}, \vol{58}, 197.

Mills, B.~Y., Turtle, A.~J., Little, A.~G., Durdin, J.~M.: 1984, \journal{Aust. J. Phys.}, \vol{37}, 321.

Payne, J.~L., White, G.~L., Filipovi{\'c}, M.~D.: 2008, \MNRAS, \vol{383}, 1175.

Ridley J. P., Lorimer D. R., 2010, \MNRAS, \vol{406}, L80.

Sault, R. J., Teuben, P. J., Wright, M. C. H.: 1995, in ``Astronomical Society of the Pacific Conference Series, Vol. 77, Astronomical Data Analysis Software and Systems IV'', R. A. Shaw, H. E. Payne, \& J. J. E. Hayes, ed., 433.

Sault, R.~J., Wieringa, M.~H.: 1994, \journal{Astron. Astrophys. Suppl. Series}, \vol{108}, 585.

Smartt S. J., 2009, ARA\&A, \vol{47}, 63.

Smith, C., Points, S., Winkler, P. F.: 2006, \journal{NOAO Newsletter}, \vol{85}, 6.

\endreferences

}

\end{multicols}

\vfill\eject

{\ }



\naslov{MULTIFREKVENCIONA RADIO POSMATRA{NJ}A OSTATKA SUPERNOVE U VELIKOM MAGELANOVOM OBLAKU -- {\bf \SNR (N59B)}} 



\authors{L.M. Bozzetto$^1$, M.D.~Filipovi\'c$^1$, E.J. Crawford$^1$, A.Y. De Horta$^1$, M. Stupar$^{2,3}$}

\vskip3mm


\address{$^1$School of Computing and Mathematics, University of Western 
Sydney
\break Locked Bag 1797, Penrith South DC, NSW 1797, Australia}  
\address{$^2$Department of Physics, Macquarie University, Sydney, NSW 2109, Australia}
\address{$^3$Australian Astronomical Observatory, PO Box 296, Epping, NSW 1710, Australia}

\Email{m.filipovic@uws.edu.au}

\vskip3mm


\centerline{\rrm UDK \udc}

\vskip1mm

\centerline{\rit Originalni nauqni rad}

\vskip.7cm

\begin{multicols}{2}

{


\rrm 
U ovoj studiji predstav{lj}amo nove {\rm ATCA} rezultate posmatra{nj}a u radio-kontinumu ostatka supernove u Velikom Magelanovom Oblaku -- \textrm{\SNR}. Ovaj objekat je tipiqan ostatak supernove sa {lj}uskastom morfologijom. Izmerena vrednost dijametra iznosi \mbox{{\rm D=(36$\times$29)$\pm$1}} parseka. Na severnoj strani ovog ostatka supernove vidljiv je jak {\rm H\,II} region koji dodatno ote{\zz}ava precizno mere{nj}e (npr. gustine fluksa i polarizacije) i da{lj}u deta{lj}niju analizu ovog objekta. Radio-kontinum emisija je identiqne strukture kao i na ostalim frekvencijama (optiqka i rentgenska) a povr{\ss}inski sjaj ovog ostatka na {\rm 1~GHz} je proce{nj}en na {\rm 2.55$\times$10$^{-21}$ Wm$^{-2}$ Hz$^{-1}$ sr$^{-1}$}. Konaqno, otkrili smo i distinktivni taqkasti radio-kontinum objekat koji je najverovatnije tesno povezan sa prethodno predlo{\zz}enim pulsarom i {\rm PWN}. Takozvani "rep" ovog ostatka nije vi{dj}iv ni na jednom od na{\ss}ih radio mapa iako je veoma prominentan i na optiqkim i na rentgenskim frekvencijama.



}\end{multicols}

\end{document}